# Programmable Optical Spectrum Shapers as Computing Primitives for Accelerating Convolutional Neural Networks

Georgios Moustakas, Adonis Bogris, *fellow member Optica*, Charis Mesaritakis

*Abstract*— **Photonic convolutional accelerators have emerged as low-energy alternatives to power-demanding digital convolutional neural networks, though they often face limitations in scalability. In this work, we introduce a convolutional photonic accelerator that employs programmable kernels manifesting as trainable waveforms in the frequency domain to enable low-energy, high-throughput scalable image classification. The proposed scheme inherently provides dimensionality reduction and feature extraction directly in the optical domain. Numerical results targeting the Fashion-MNIST show that by using only 16 optical nodes, the system's classification accuracy tops at 90.1%, when typical backpropagation is used. Moreover, by adapting the training technique to forward-forward approach a marginal drop by 1% is recorded compared to the back-propagation scenario, thus showcasing the compatibility of the overall architecture with a hardware friendly training approach. Finally, we experimentally implement the trained kernels using a programmable waveshaper. Despite the difference between the simulated and experimentally generated transfer functions of the programmable Kernels, the classification accuracy based on the experimentally obtained kernels exhibits a marginal 0.2% reduction proving the validity of the idea and its high robustness to variations of the frequency applied complex weights.**

## I. INTRODUCTION

ARTIFICIAL intelligence has evolved from an early conceptual attempt to emulate human intelligence into one of the most promising and widely used technologies of the modern era, enabling a wide range of applications including computer vision, speech recognition, natural language processing, and biomedical research. One of the earliest milestones in AI, particularly in computer vision, was LeNet [1] , developed by LeCun *et al.* in the late 1990s. LeNet demonstrated that convolutional neural networks (CNNs) could successfully classify handwritten digits from the MNIST dataset. By exploiting the convolution operation and its properties, it enabled CNNs to process images efficiently, significantly reducing the computational overhead associated with fully connected networks, which require a separate weight for each pixel. However, due to the lack of computational resources at that time, the widespread adoption of deep networks was delayed until the introduction of the multilayer convolutional network AlexNet [2], which was trained on the ImageNet dataset [3] using a graphics processing unit (GPU). This breakthrough reignited research in the field and led to the development of more advanced and complex CNN architectures, such as InceptionNet [4], ResNet [5], and numerous successors. While deep CNN architectures are able to increase classification accuracy it is known that due to their deep structure the number of trainable parameters and corresponding operations skyrockets along with the overall power consumption. More specifically, it has been demonstrated that in convolutional layers the multiply and accumulate (MAC) operations account for over 99% of the total operations in state-of-the art CNN architectures, thereby significantly impacting energy consumption [6]. Recent efforts seek to mitigate the trade-off between accuracy and energy consumption through alternative, analogue approaches concerning network design and training strategies [7]. Integrated photonics is one of the most prominent platforms for disrupting the machine-learning (ML) field, through their low power consumption, massive parallelism and low-latency signal processing. Based on these merits, in the recent years, there has been intensive research in photonic convolutional neural network accelerators for image processing and classification tasks. More specifically, photonic CNN (PCNN) architectures can be classified into three broad categories based on whether information is encoded and processed in the temporal, spatial, or spectral domain.

Starting with the spatial class of photonic CNN networks, in such systems the 2D data images are physically imprinted onto the beam of light or the wavefront, with each pixel's intensity directly mapped to the light's amplitude (or phase) at different $x,y$ coordinates. The convolution is not performed through a series of multiply-and-accumulate operations but is

This work was supported by the Research Project QUASAR which is implemented in the Framework of H. F. R. I Call "Basic Research Financing (Horizontal Support of All Sciences)" under the National Recovery and Resilience Plan "Greece 2.0" funded by European Union—NextGenerationEU under Project 016594 and the EU Horizon Europe PROMETHEUS project under grant agreement 101070195.

Georgios Moustakas is with the University of West Attica, Department of Informatics and Computer Engineering Egaleo, Agiou Spiridonos (e-mail: gemoustakas@uniwa.gr).

Adonis Bogris is with the University of West Attica, Department of Informatics and Computer Engineering Egaleo, Agiou Spiridonos (e-mail: abogris@uniwa.gr).

Charis Mesaritakis is with the University of West Attica, Department of Biomedical Engineering, Egaleo, Agiou Spiridonos (e-mail: cmesar@uniwa.gr).

instead executed as the light propagates through standard Fourier optics. Experimental results have demonstrated classification accuracies of up to 93% on MNIST dataset, 87.5% on Fashion MNIST [8] and while another work has demonstrated up to 44.4% on the CIFAR10 dataset [9]. Such systems often rely on free-space optical components, spatial light modulators (SLMs), and digital micromirror devices (DMDs), which are typically bulky and not fit for integration. Very recently space-efficient optical convolutional processors based on integrated chip diffractive neural schemes have been demonstrated with classification accuracy of 80% for the Fashion MNIST [10].

Temporal-domain-based PCNNs handle 2D image data in a fundamentally different way. Specifically, the 2D image is flattened and treated as a 1D vector, with information encoded in time rather than in space. The serialization process is carried out in the electrical domain and the data are typically encoded onto the amplitude of a continuous-wave (CW) laser. Multi-wavelength approaches are followed in order to imprint the Kernel weights. For instance, for a 4x4 Kernel realization, 16 wavelengths will be incorporated. Stride information is also included in the 1D vector leading to high representational redundancy. Experimental results have demonstrated a classification accuracy of approximately 90% on the MNIST dataset [11],[12],[13].

An elegant proposition to photonic CNN is the use of interference of coherent light. Mach-Zehnder Interferometers in a mesh configuration constitute a proper platform for matrix vector multiplications [14]. The majority of the previously mentioned photonic solutions aim to speed-up the matrix-vector-multiplications (MVM) that are required to execute digital convolution operations by transferring them to the analogue domain. However, such an approach comes with several disadvantages in terms of scalability, as for larger images or networks, the number of photonic components or resources (such as wavelengths) is increased generating stringent constraints in terms of footprint and power consumption [15].

Training approach of optical CNNs remains a central issue. While in conventional digital CNNs, backpropagation is the standard approach, physical systems demand a different methodology; primarily because such systems require precise characterization prior to training, whereas a differentiable function is not always straight-forward to find [16]. For such applications, the forward–forward algorithm, proposed by Hinton et al. [17] has been applied to train various physical systems, including photonic neural networks [8], [16], [18]. The forward–forward algorithm is advantageous because it computes a local loss function, eliminating the need to backpropagate a loss signal through each layer, allowing the independent training of each layer separately, as a black box.

In this work, we propose a novel scheme to optical CNNs which seeks to propose a practical and powerful approach addressing significant issues of state-of-the-art solutions mainly relying on photonic integrated approaches offering MVM operations for convolutional processing. We build upon the optical spectrum-slicing approach of Tsirigotis et al. [19][20], which performs convolutions through the application of multiple optical filters, which are detuned compared to the central frequency of the signal and have limited bandwidth leading to spectral slicing of the incoming signal. Whilst this technique is adequate to boost classification accuracy compared to standalone FCLs, it can not reach the performance of digital implementations due to its limited capabilities in terms of training and adaptation. Here, we further expand this work by introducing convolutional processing employing arbitrarily programmable optical filters (POF-CNN). POF-CNN uses fully trainable analogue kernels in the frequency domain through fine tuning the spectral response of filter nodes. By assuming realistic spectral resolution values in the order of a few GHz, we propose a lightweight and low-power image classification system. In the proposed technique, contrary to previous MVM-based accelerators, the number of POF-CNN nodes is independent of the dataset size and no severe data expansion is required in 1D flattening. Additionally, single wavelength operation suffices and the dimensionality reduction provided by the photodiode at the end of each node prior to analog-to-digital conversion (ADC) reduces data volume, power consumption, and ADC constraints on sampling rate. Combined with the encoding of image data into the optical signal, this allows convolution to be executed in a single optical pass. In order to propose a holistic approach regarding inference and training, we adopted both conventional backpropagation and the forward-forward algorithm to train our scheme, with the latter offering a more practical route for in-situ training. When trained with backpropagation, accuracy tops at 90.1% assuming 16 programmable optical filter-nodes, while a more lightweight scheme of 6 filters reaches 89.4%. Forward-forward algorithm yields an accuracy of 88.54% for the same lightweight configuration establishing a practical route to in-situ training. In order to verify the feasibility of the idea, we generated the transfer functions of the trained Kernels with the use of a commercially available waveshaper and estimated the declinations when experimental filter coefficients are used in the scheme, subject to resolution and accuracy constraints. The experimentally obtained transfer functions offer a classification accuracy reduced by 0.2% relative to the ideal case proving the validity of the idea and the robustness of the system to filter parameter variations.

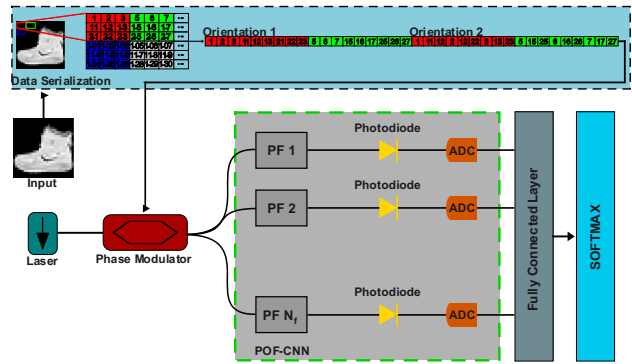

Figure 1: Schematic representation of the proposed architecture

The rest of this paper is organized as follows: Section I introduces the proposed concept and architecture; Section II discusses the numerical methods that were followed; Section

III presents the simulation results; and lastly, Section IV presents a potential practical implementation along with the work's conclusion.

## II. CONCEPT

In this section, we introduce POF-CNN which constitutes a passive all-optical convolutional accelerator that relies on spectrally trained optical programmable filters to enable low-power and lightweight image processing through convolution operations. In contrast to what is usually followed in digital CNNs, instead of considering trainable convolutional Kernels in the spatial-temporal domain, in this work the filters are considered programmable in the frequency domain, which is mathematically and physically equivalent. The 2D convolution, without nonlinearity and bias term is expressed as (1):

$$y[c_{out}, i, j] = \sum_{c_{in}=0}^{c_{in}-1} \sum_{u=0}^{U-1} \sum_{v=0}^{V-1} X[c_{in}, i-u, j-v] \cdot W[c_{out}, c_{in}, u, v]$$

Where $c_{out}, c_{in}$ are the output, input channels respectively and $U \times V$ is the kernel size. The following equation can be reduced to 1D form by setting V = 1, dropping index j and by focusing into 1 input/output channel ($c_{out} = c_{in} = 1$), equation (1) is reduced into typical discrete time convolution operation as (2):

$$y[n] = \sum_{k=0}^{K-1} X[n] \cdot W[n-k]$$

Which in the analogue domain, is equivalent to (3):

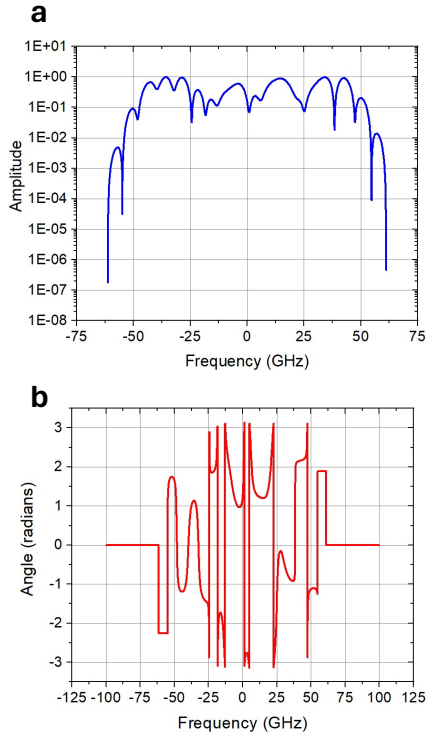

Figure 2: (a) Amplitude (b) Phase response of an arbitrary filter with 6 GHz resolution

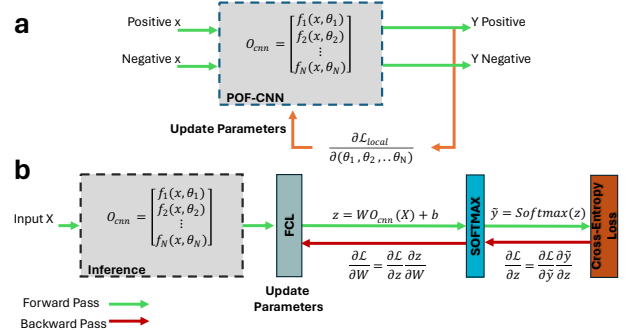

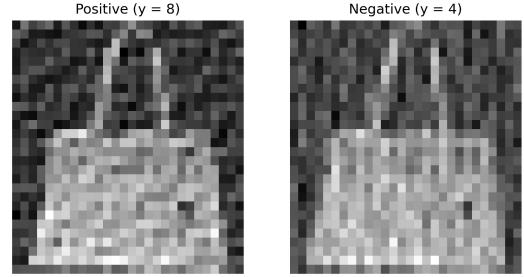

Figure 3. (a) POF-CNN local training (b) Forward-Forward training based on goodness + cross entropy loss

Positive (y = 8)    Negative (y = 4)

Figure 4. Input image overlaid with correct (left) and incorrect (right) label

$$y(t) = \int x(t) \cdot w(t-\tau) \, d\tau$$

Using the convolution property of Fourier transform, the output can be calculated in frequency domain as $Y(f) = X(f) \cdot W(f)$ where $W(f)$ is the transfer function of the filter, and the response in time domain can be acquired using the inverse Fourier transform. Instead of using a training algorithm to identify $w(t)$ in the temporal domain, one can equivalently compute its transfer function $W(f)$ in the spectral domain. Based on available photonic technologies, the filters can be approximated as a discrete set of $N_p$ equally spaced complex points in the frequency domain. If one wants to cover the entire optical bandwidth of the signal to be processed defined as $BW_{sig}$, then the resolution of the filter approximation is expressed as $\Delta f_{res} = \frac{BW_{sig}}{N_p}$. Hence, the resolution of an arbitrary waveform generator in the frequency domain, for instance the resolution of a waveshaper, will define the accuracy in determining the programmable transfer function and also will regulate the number of trainable parameters for the POF-CNN. Typical commercially available waveshapers can offer resolution in the order of a few GHz, whilst recent research papers show resolution below 1 GHz [21]. Taking into consideration that state of the art digital kernels for CNNs have a dimension of 7x7 (49 points) [22] the 1 GHz spectral resolution alongside the availability of electro-optic modulators with bandwidth of 50GHz, results that the proposed scheme can offer an equivalent photonic kernel with equal number of free parameters compared to digital CNNs. For CNN processing, the aim is to train these points and identify their optimal values using methods such as backpropagation (BP) [23] and gradient descent in order to learn the most suitable transfer function for the given task.

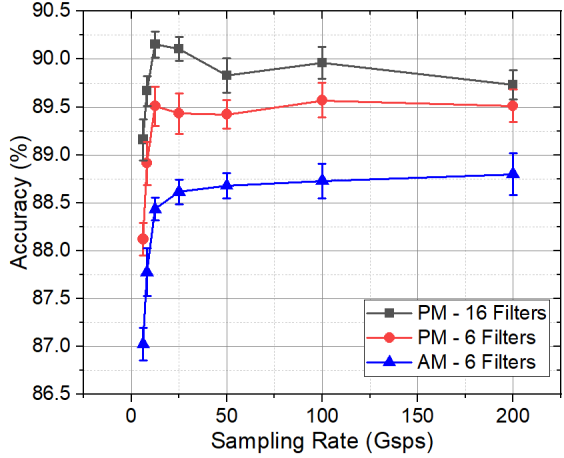

Figure 5. Classification accuracy as a function of sampling rate

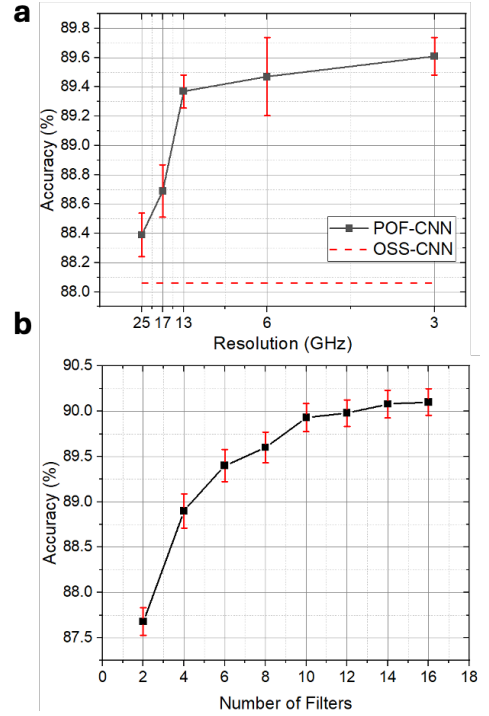

Figure 6: Classification accuracy vs (a) filter resolution ($\Delta f_{res}$) (b) number of filters ($N_f$)

In [19][20] the authors have shown that convolutional processing is benefited by optical filtering approaches in the form of spectrum slicing with the use of typical bandpass filters. In these papers, the optical pre-processing scheme is based on fixed filtering units in terms of their spectral shape and the only hyperparameters that were investigated are the detuning between the different filters contributing to the slicing process and their 3 dB bandwidth. Here, we generalize this idea introducing a refined learning mechanism that enables more precise parameter tuning; instead of using well-defined filters (e.g Gaussian filters) as in [19], the overall transfer function covering the entire bandwidth is learned through an iterative data-driven process based on $\Delta f_{res}$ and $N_p$ parameters. The amplitude and phase response of a such an arbitrary transfer function is illustrated in Fig. 2 with $N_p$=16 trainable points yielding a resolution of about $\Delta f_{res}$=6 GHz considering $BW_{sig}$= 100 GHz. The system architecture is depicted in Fig. 1, where as in [19], each image is divided into 4 x 4 patches that are serialized into a single vector containing all initial pixel values (intensities) flattened with two orientations, namely row and column major order. Compared to techniques in [11], [13] that perform 1D flattening with stride=1 and Kernel size=4x4, the amount of data and thus the resulted latency is vastly reduced. For instance, for a MNIST image (28x28) (no padding), the number of extracted patches is 625 with each of them corresponding to 4x4 pixel vectors which results in a total flattened dimensionality of 10000 elements. In our case, no stride is required and the total flattened dimensionality is equal to 2x(28x28)=1568 elements These orientations are chosen to better map the 2D spatio-temporal correlations of the initial image to the flattened 1D vector permitting only temporal correlations at a moderate redundancy. An optical modulator superposes the image vector values onto the phase or the amplitude of a continuous-wave optical carrier in a sequential manner. The photonic convolutional layer consists of $N_f$ kernels (denoted as PF in the graph) followed by a photodiode and an ADC. After training, the filters will have distinct transfer functions. Through the convolution of their impulse responses with the input, they will be able to extract different features, as in traditional CNNs. The non-linearity in this system is provided by each photodiode that detects the time-traces of the filters and at the same time performs the averaging operation on the convolved data similar to an AvgPooling layer. Another nonlinearity factor is the phase to intensity conversion and its transformation by the distinct kernels in the case that phase modulation is utilized. Since the phase to intensity nonlinearity depends on the transfer function of the filter, each filter provides a different nonlinear transformation to the phase modulated input. Following the photodiode, an ADC is used to convert the analog time series into digital samples. Finally, the resulting data are serialized using a flattening layer and fed into a digital back-end consisting of a fully connected layer for classification.

## III. NUMERICAL METHODS

This section presents all the numerical methods employed in this work to simulate the programmable filter operation in a realistic way. The most critical hyperparameters of the convolutional accelerator are the filter resolution $\Delta f_{res}$ which is inverse proportional to the number of trainable points $N_p$, the number of filters (kernels) $N_f$ and photodiode bandwidth which determines ADC sampling rate. The hyperparameter tuning and classification performance evaluation is conducted using the Fashion-MNIST [24] dataset of fashion garments, which consists of 60,000 images for training and 10,000 images for testing. Each image is vectorized according to a 4 x 4 patch where each patch is serialized with row and column major orientation denoted as orientation A and B respectively. The resulting 2 x (28 x 28) vectors drive a simulated Mach-Zehnder modulator used to encode the vector values onto the phase or amplitude of a CW carrier. In case of phase modulation, the input signal is $E_{in} = \sqrt{P}e^{i*m(t)}$ where $m(t)$ is the input vector with minimum value equal to 0 and maximum value equal to $\pi$ and $P$ is the mean optical power in mW. Amplitude modulation was also investigated with a maximum modulation depth of 90%. Then the signal is split to the $N_f$ filters whose amplitude and phase values across the $N_p$ points are randomly selected using Xavier (He) Initialization [25]. Then as specified in concept, on each forward pass the output of each filter is calculated in the frequency domain as $Y_i(f) = X(f) \cdot H_i(f)$ where $H_i(f)$ is the transfer function of the i-th filter. In order to emulate analogue processing in the frequency domain, we approximate the analogue transfer function of the i-th filter through spline interpolation of $N_p$ points. The output signal is calculated as $y_i(t) = FFT^{-1}\{Y_i(f)\}$. The resulting electric field is driven to a photodiode simulated as a square law detector affected by thermal and shot noise with a noise equivalent power equal to $25\ pW/\sqrt{Hz}$, followed by a fourth order Butterworth filter which mimics its low-pass frequency response [20]. The parameters that determine the performance in terms of accuracy and complexity are the $N_p$ points connected with frequency resolution $\Delta f_{res}$, $N_f$, $BW_{PD}$ which affects averaging and final sampling rate $f_s$=2.5 $BW_{PD}$. $N_p$ and $N_f$ affect the complexity of the training process of the photonic part whilst $N_f$, $f_s$ define the number of parameters that will feed the digital part of the network. The fully connected back-end can grow significantly in terms of trainable parameters, especially at high sampling rates and/or with a large number of filters. To reduce the high parameter count, weight pruning was employed using the L2 norm as a metric to identify filters that contribute the least to the overall classification accuracy. Due to the end-to-end differentiability of all operations within the specified pipeline, the backward pass enables gradients to propagate from the loss function all the way to the kernel points $N_p$. This allows gradient-based optimization algorithms, such as stochastic gradient descent, to iteratively adjust the kernel parameters in a manner that minimizes the loss and improves model performance.

Apart from typical back-propagation, forward-forward (FF) training based on Hintons' pioneering work is also investigated. The FF training is used to train the photonic part of the neural network with the goodness function defined by Hinton, where the aim of the learning is to make the goodness be well above some threshold for the positive/real data and well below the threshold for the negative/perturbed/distorted data [17]. We evaluate our photonic accelerator using the Forward–Forward (FF) algorithm, as it is widely recognized as a suitable learning mechanism for in-situ training. FF requires only forward passes—eliminating backward gradient propagation—and employs local learning rules, thereby minimizing memory requirements. Afterwards, a linear classifier is trained on the activations of the photonic CNN using the cross-entropy loss function and BP. As shown in Fig. 3, negative and positive samples are generated from input tensors by overlaying the labels, as proposed in [17], [26]. A local goodness function is then applied to train the POF-CNN convolutional kernels with a threshold set equal to the number of neurons as originally suggested by Hinton. Once the POF-CNN kernels are trained using the goodness function, the digital backend is trained on their activations using the standard pipeline of cross-entropy loss and backpropagation. It is important to highlight that when training the digital backend, the labels are not overlaid on the inputs, as this would cause label leakage. Instead, the POF-CNN kernels are frozen and used only for inference on raw tensors after flattening and modulation. In Hintons' original work, the labels are embedded on the input by one-hot encoding the label to the first 10 pixels but as explained in [26], this method cannot work in convolutional neural networks. In the same sense, labels are overlaid on the input by assigning each class a distinct set of sinusoidal frequency components. More specifically, after flattening, each sample is augmented in the electrical domain with a superposition of 16 orthogonal sinusoidal signals, whose frequencies are randomly sampled from a uniform grid spanning 1 to 2 kHz. These frequencies differ across class labels, giving each sample a unique frequency signature that can later be used to distinguish "good" from "bad" samples. The resulting images, overlaid with the correct and incorrect labels are shown in Fig. 4. Model training and overall system simulation were performed using PyTorch's GPU backend and automatic differentiation engine [27], running on an NVIDIA Titan GPU with 24 GB of memory. The Cross-Entropy loss function is employed in the case of BP and Hintons' goodness function in the case of FF training. Weight optimization was carried out using Adam's algorithm [28], with default β-values of 0.9 and 0.999, and a learning rate of 0.001 in the case of BP, while an epoch-dependent learning rate was used in the case of FF similar to [26]. Finally, a batch size of 64 is chosen empirically based on performance considerations, and an early stopping mechanism was employed. The patience parameter—defined as the number of epochs to wait after the last improvement in the monitored metric (e.g., validation cross-entropy loss or validation goodness)— was set to 50.

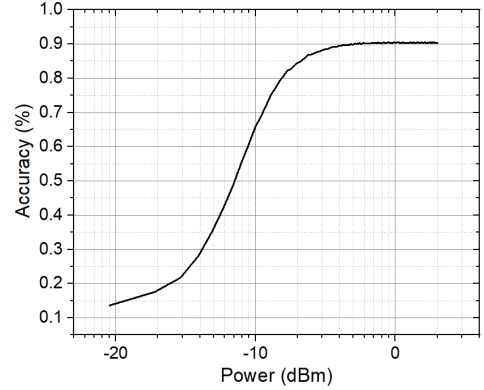

Figure 7: Classification accuracy vs launched power

After training the system, the coefficients of each filter become available and an experimental approximation of the transfer function is conducted to provide more realistic transfer functions in the simulated system. A setup consisting of diverse filters with a frequency resolution $\Delta f_{res} = 12GHz$ was experimentally evaluated, where the transfer functions are approximated using a Finisar WaveShaper A1000. More specifically after normalizing the coefficients for each filter, the corresponding attenuation values corresponding to the amplitude coefficients of the filter were computed and sequentially mapped. A tunable continuous-wave (CW) source (CoBrite DX2) was used to measure the amplitude of each filter's transfer function via a sequential frequency scan

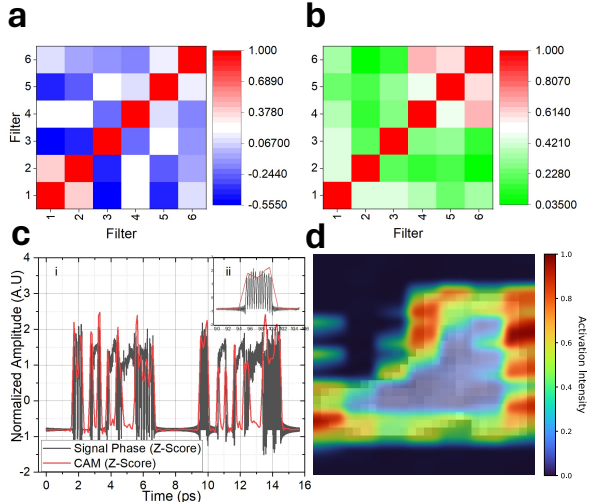

Figure 8: Cross Correlation of (a) phase (b) magnitude between POF-CNN kernels (c) i. class activation map (c) ii. zoomed in inset from 9 to 10 ps (d) Normalized 2D equivalent of (c)

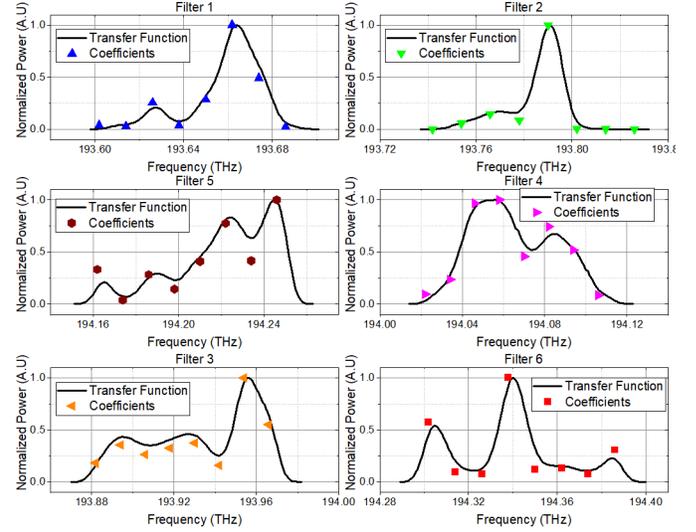

Figure 9: Experimental Transfer Function acquired by the Waveshaper (line) and Simulation Coefficients (points) that refer to the amplitude of the trained complex weights

spanning 193.50–194.50 THz, with the optical power recorded using a Thorlabs PM100D power meter for offline processing. After measuring the transfer functions, the mean squared error (MSE) between the experimental and simulated coefficients is computed. Finally, the simulated coefficients are subsequently perturbed by noise with a standard deviation defined in Eq. (5).

$$\sigma = \sqrt{\frac{MSE}{2}} \qquad (5)$$

Where MSE is the mean square error between experimental and simulated coefficients. Let the coefficients for filter $k$ being defined as:

$$c_k = c_{k,r} + j c_{k,i} \qquad (6)$$

Then, the noisy coefficients can be expressed as

$$n_k = n_{k,r} + j n_{k,i} \qquad (7)$$

with $n_{k,r}, n_{k,i} \sim \mathcal{N}(0, \sigma^2)$, then the final distorted symbols ($\widetilde{c_k}$) can be expressed as (6):

$$\widetilde{c_k} = c_k + n_k \qquad (8)$$

## IV. RESULTS AND DISCUSSION

As introduced in the methods section, different parameters such as $N_p$, $N_f$, $BW_{PD}$ affect the complexity and accuracy of the POF-CNN. We first focus on the effect of $BW_{PD}$ which determines the analog average pooling performed at the photodetector level. The results are demonstrated in Fig. 5 on two classification systems consisting of $N_f$=6 and 16 filters respectively, each with filter resolution $\Delta f_{res}$=6 GHz ($N_p$=16) to process a 100 Gpixels/sec signal using phase modulation and a single system with $\Delta f_{res}$=6 GHz ($N_p$=16) and $N_f$=6 using amplitude modulation. The figure shows that in both filter configurations, classification accuracy increases with higher sampling rates, but begins to saturate beyond 12.5 GS/s. This saturation highlights the trade-off between dimensionality reduction and lossy data compression. The dimensionality reduction offered by the low-pass filtering process exhibits its optimum value at $BW_{PD}$=5 GHz ($f_s$=2.5 $BW_{PD}$) for a 100 Gpixels/sec signal as discussed above. To simplify the system and reduce the number of samples fed into the linear backend, this rate was selected in the simulations, as the resulting accuracy is comparable to that of a more complex setup using a 200 Gsps ADC. It must be noted that the accuracy performance achieved for FMNIST dataset is very high (~90%) and approaches or even surpasses that offered by established digital CNN models. A comparison analysis is included in this section below. As depicted in Fig. 5, when comparing the two configurations using phase and amplitude modulation (both using $N_f$=6) the phase modulation variant consistently achieves higher classification accuracy which probably relates to the enhanced nonlinear phase to intensity features offered by phase modulation. For this reason, all the results that follow consider solely phase modulation.

Fig. 6(a) illustrates the classification accuracy as a function of filter resolution $\Delta f_{res}$. The X-axis represents the filter resolution in GHz in a classification system that uses $BW_{PD}$=5 GHz, $f_s$=12.5 GSps, $N_f$=6 for the photonic accelerator layer and 100 Gpixels/sec input signal rate which corresponds to 100 GHz signal bandwidth. As depicted, to maximize the classification accuracy, a filter resolution of at least $\Delta f_{res}$=6 GHz ($N_p$ = 16) is required for a signal bandwidth of 100 GHz. Such a resolution value is practical and already offered by commercially available products. A comparison with the OSS technique first proposed by Tsirigotis et al. is provided, on a classification setup comprising of 6 filters with 25 GHz bandwidth. Figure 6 demonstrates that POF-CNN significantly outperforms OSS-CNN as a result of its better resolution in defining weights in the spectral domain and its optimization through established learning techniques such as BP. It is important to note that further increasing the number of trainable points enhances the frequency resolution, resulting in finer spacing between samples or points and potentially improving overall classification accuracy, but it may also pose challenges for practical implementation. This introduces a trade-off between feasibility and the number of trainable points and for the considered signal bandwidth $BW_{sig}$=100 GHz, a resolution of $\Delta f_{res}$=6 GHz is adequate to ensure high classification accuracy (~89.5% in Fasion MNIST) and practicality. Figure 6(b) shows the classification accuracy as a function of the number of filters in the photonic accelerator layer, for $BW_{PD}$=5 GHz photodiode and $\Delta f_{res}$= 6 GHz ($N_p$=16). The results indicate that classification accuracy improves with an increasing number of filters, reaching a plateau beyond 12 filters where accuracy surpasses 90% for FMNIST. Further increasing the number of filters might not provide any benefit; on the contrary it can lead to overfitting and complicated practical implementation.

Classification accuracy as a function of launched power is demonstrated in figure 7 for a classification setup that consists of $N_f$=6 filters, $N_p$=16 trainable points and $BW_{PD}$=5 GHz. It is evident that when the injected power is substantially low, the noise from the photodiode, most notably thermal noise, significantly impacts the classification accuracy. However, as the injected power increases, the signal-to-noise ratio (SNR) improves, and the effect of PD noise on classification accuracy becomes negligible. The performance seems to be stabilized for input power higher than -8 dBm for the considered $BW_{PD}$.

Extended simulation runs were performed using $BW_{PD}$=5 GHz considering two moderate architectures using $\Delta f_{res}$=6.25 resolution ($N_p = 16$) with $N_f$=6 in the convolutional layer and a more demanding architecture of $N_f$=16 in the convolutional layer. Table I summarizes the classification accuracy of the proposed architectures in comparison to a single layer CNN comprising of 6 filters each with a 4x4 kernel followed by Average Pooling Layer and ReLU non-linearity, LeNet-V5 and ResNet-18. The number of trainable parameters for POF-CNN includes the $N_f * N_p$ parameters of the programmable kernels, but it is predominantly determined by the size of the linear backend, which—as stated above—scales with $N_f$ and with the $BW_{PD}$. More specifically, the number of parameters for the POF-CNN layer can be calculated as:

$$N_{POF-CNN} = N_{linear} + N_K \qquad (9)$$

Where the number of linear parameters is calculated as:

$$N_{linear} = (N_f \cdot N_{out} \cdot 10) + 10 \qquad (10)$$

$N_{out}$ are the output features of each filter after ADC downsampling, while, the number of trainable parameters for POF-CNN kernels are

$$N_K = N_p \cdot N_f \qquad (11)$$

The analysis of Table I demonstrates that ResNet-18 is able to reach up to 93.57% classification accuracy but with 11,175,370 trainable parameters, LeNet-V5 is not able to surpass 90% accuracy, reaching 89.59% with 44.470 trainable parameters while single layer CNN and FCN remain below 89% accuracy with 8K and 7K trainable parameters. The POF-CNN architecture achieves over 89% accuracy, with 89.41% obtained with $N_f$=6 and $N_p$=16, requiring only 11,866 trainable parameters. Accuracy increases to 90.1% in the case of $N_f$=16 filters with the same filter resolution, which uses three times the trainable parameters. This demonstrates that even with a more computationally demanding configuration of 16 filters but with the same resolution, the accuracy surpasses that of LeNetV5, while using 13K fewer trainable parameters. It should be noted that in practical/experimental scenarios, the kernels are implemented in the analogue domain. Consequently, the number of trainable parameters is practically determined/affected by the number of trainable parameters in the linear backend exclusively, further reducing the complexity.

Table I : Classification and number of trainable Parameters for different architectures

| Network | Classification Accuracy (%) | # Trainable Parameters |
|---|---|---|
| Single layer CNN | 88.1 | 8,752 |
| LeNetV5 | 89.59 | 44,470 |
| ResNet18 | 93.57 | 11,175,370 |
| POF-CNN_6-6.25GHz | 89.41 | 11,866 |
| POF-CNN_16-6.25GHz | 90.11 | 31,626 |

Table II : Classification and number of trainable parameters for various pruned networks

| # Filters | Classification Accuracy (%) | # Trainable Parameters |
|---|---|---|
| Baseline (16) | 90.11 | 31,626 |
| 15 | 89.73 | 29,666 |
| 14 | 88.67 | 27,706 |
| 12 | 88.65 | 23,786 |
| 11 | 85.89 | 21,826 |
| 10 | 82.65 | 19,866 |

Pruning results for a system of $N_f$=16 and $N_p$=16 resolution are shown in Table II, demonstrating that complexity of the linear layer can be reduced without a significant drop in classification accuracy when up to four filters are removed. Pruning beyond this point leads to a noticeable reduction in accuracy compared to the baseline (16 filters), mainly because it removes filters that extract features critical for distinguishing between classes.

Figure 8 shows the cross-correlation between the phase and magnitude responses of six POF-CNN kernels. The results indicate a mean correlation of 0.324 for magnitude and −0.053 for phase, highlighting that each kernel successfully encodes distinct features. Finally, a 1D equivalent of the Gradient-weighted Class Activation Mapping (Grad-CAM) [29] was obtained to highlight the temporal regions of the modulated signal that contribute most to the label prediction. First, a global average of the gradients is calculated and used to compute a weighted sum of the activations. Unlike standard digital CNNs, which typically employ ReLU at this stage, the absolute value of the weighted sum is used here. Figure 9c demonstrates the Grad-CAM output for six POF-CNN kernels, showing that they act as edge detectors, as indicated by increased activations along the edges of the image. This suggests that the kernels effectively extract features associated with rapid variations in the phase of the modulated signal. This behavior is more clearly illustrated in Figure 9d, which presents the two-dimensional equivalent, where strong activations are concentrated at the edges, while the smoothly varying regions of the image exhibit suppressed responses.

Table III : Classification Accuracy for FF and BP

| Network | Classification Accuracy (%) | Standard Deviation | Maximum Accuracy (%) |
|---|---|---|---|
| **POF-CNN – Backpropagation** | 89.41 | 0.1741 | 89.67 |
| **POF-CNN – Forward Forward** | 88.54 | 0.004 | 89.25 |

Since the analog part of POF-CNN is not easy to train in-situ using back-propagation, we also consider FF training as documented in the methods. The results of FF training on a setup comprising of $N_f$=6 filters with $N_p$=16 and $BW_{res}$=5 GHz are summarized in table III where it is shown that FF achieved a mean classification accuracy of 88.54% across trials, with a standard deviation of 0.004 and a maximum accuracy of 89.25%, showing a performance degradation in mean classification accuracy of approximately 0.8% when compared to the same system trained with the backpropagation algorithm. This result establishes a foundation for enabling in-situ training of the POF-CNN.

The applicability of the POF-CNN in real experimental conditions is studied with the emulation of the filters' transfer functions as provided by back-propagation method with the use of a commercially available waveshaper. 6 filters each with $\Delta f_{res} = 12$ GHz were approximated using the waveshaper with the resulting normalized power values along with the amplitudes of the coefficients demonstrated in Fig. 9. The experimentally measured transfer functions closely match the simulated coefficients with an average mean squared error of 0.6%. To examine the impact of deviations between the experimental and simulated coefficients on classification accuracy, deviations in the form of noise were introduced into the coefficients (as discussed in detail in the Methods section), and inference was performed on the test dataset. The results show that the baseline network achieved a classification accuracy of 89.20%, while the network with distorted coefficients reached 88.98%, indicating a slight degradation of approximately 0.22% in accuracy providing the high robustness of the specific CNN approach in noise or other instabilities.

The proposed architecture, being a purely analog and non-conventional computing approach, cannot be directly compared to digital processing units such as GPUs in terms of operations per second. Instead, its performance is primarily limited by the throughput of the arbitrary waveform generator used to superpose the data onto the modulator, as well as by the input/output latency, which remains very low for single-pass processing. It is important to note that training with either back-propagation or forward-forward learning methods achieves optimal performance with a moderate number of epochs—typically fewer than 250—which is comparable to the training requirements of the digitally implemented CNNs considered in this work. Notably, the POF-CNN attains similar or even superior classification accuracy compared to multi-layer CNNs such as LeNet, which incorporates two convolutional layers. This demonstrates that the single-layer CNN architecture of the POF-CNN is both powerful and well suited for accelerating tasks of moderate complexity. Future work will focus on extending the POF-CNN framework to support multi-layer operation.

## IV. CONCLUSION

This work proposes a programmable photonic analog convolutional accelerator that operates in the spectral domain. Through a single optical pass, the proposed scheme performs convolution using programmable spectral-domain filters with minimal electro-optic conversions, while nonlinear transformations are implemented via phase modulation and photodiode nonlinearity. Pooling is carried out directly in the analog domain, and dimensionality reduction is achieved using reduced-sampling-rate ADCs. Training of the kernels was performed using both standard backpropagation and the forward-forward algorithm. Numerical results demonstrate classification accuracies of up to 90.11% on the Fashion-MNIST dataset using backpropagation and 88.54% when using the forward-forward algorithm which match those offered by state of the art digital CNN models. Experimental realization of the kernels showed only ~0.2% degradation in performance compared to noise free simulation. By leveraging recent advances in programmable filters within silicon photonic technologies , the minimal architecture of the POF-CNN processor, which requires only an optical source, a phase modulator, a few photodetectors, and low-speed acquisition electronics relative to the data rate (12.5 Gsps for 100 Gbaud signals), becomes highly attractive for enabling a low-cost and energy-efficient photonic convolutional processor.